\documentclass[a4paper,twocolumn,superscriptaddress,10pt]{quantumarticle}
\pdfoutput=1
\usepackage[utf8]{inputenc}
\usepackage[english]{babel}
\usepackage[T1]{fontenc}
\usepackage{amsmath}
\usepackage{hyperref} 
\usepackage{graphicx}
\usepackage{verbatim}
\usepackage{tikz}
\usepackage{lipsum}
\usepackage{verbatim}
\usepackage[numbers,sort&compress]{natbib}

\newcommand{\ra}{\rangle}
\newcommand{\la}{\langle}

\begin{document}

\title{Generating remote entanglement via disentangling operations}
\date{\today}

\author{Jennifer R. Glick}
\email{patte399@msu.edu}
\affiliation{Department of Physics and Astronomy, Michigan State University, East Lansing, MI 48824 USA}

\author{Christoph Adami}
\email{adami@msu.edu}
\affiliation{Department of Physics and Astronomy, Michigan State University, East Lansing, MI 48824 USA}

\maketitle

\begin{abstract}
 Shared entanglement between spatially separated systems is an essential resource for quantum information processing including long-distance quantum cryptography and teleportation. While purification protocols for mixed distributed entangled quantum states exist, it is not clear how to optimally distribute entanglement to remote locations. Here, we describe a deterministic protocol for generating a maximally entangled state between remote locations that only uses local operations on qubits, and requires no classical communication between the separated parties. The procedure may provide protection from decoherence before the entanglement is ``activated,'' and could be useful for quantum key distribution. 
\end{abstract}

\section{Introduction}

 Much of quantum information processing relies on entanglement as a resource. For example, entanglement that is shared between distant parties is necessary to implement Ekert's quantum key distribution protocol for secure communication~\cite{Ekert1991}, to transfer quantum states using teleportation~\cite{Bennett1993}, or to establish large-scale quantum networks. Remote entanglement generation has been realized in many systems such as with optical photons~\cite{Chou2005,Hofmann2012,Moehring2007}, the nitrogen vacancy centers of solid state qubits~\cite{Bernien2013}, and superconducting qubits~\cite{Roch2014,Narla2016,Koshino2017}.
 
 When manipulating entanglement, it is sometimes necessary to {\em disentangle} subsystems of an entangled system. Disentangling is, generally speaking, only possible under certain conditions~\cite{MorTerno1999}, principally because of the relation between disentangling and quantum cloning (a single quantum cannot be disentangled~\cite{Mor1999}). It is possible, however, to disentangle known states, and in particular it is possible to disentangle subsystems of jointly entangled systems as long as some information about the quantum state is obtained. 
 
 Here, we describe a method for deterministically generating remote entanglement between two qubits that have never interacted using local operations on {\em pairs} of separated qubits. We present first in Sec.~\ref{sec:LOCC} a general scheme that relies on classical communication between the remote parties, where the degree of entanglement created can be tuned by the choice of encoding parameters. From this construction, we consider in Sec.~\ref{sec:LO} a special case where the communication requirement is eliminated and the resulting entanglement between the remote parties is maximal. Unlike standard entanglement swapping protocols~\cite{Zukowski1993}, this scheme generates remote entanglement without joint measurements and, in the special case discussed in Sec.~\ref{sec:LO}, utilizes only operations applied locally to the qubit pairs.

 \section{Encoding scheme}
\begin{figure}[htbp]
   \centering
   \includegraphics[width=\linewidth]{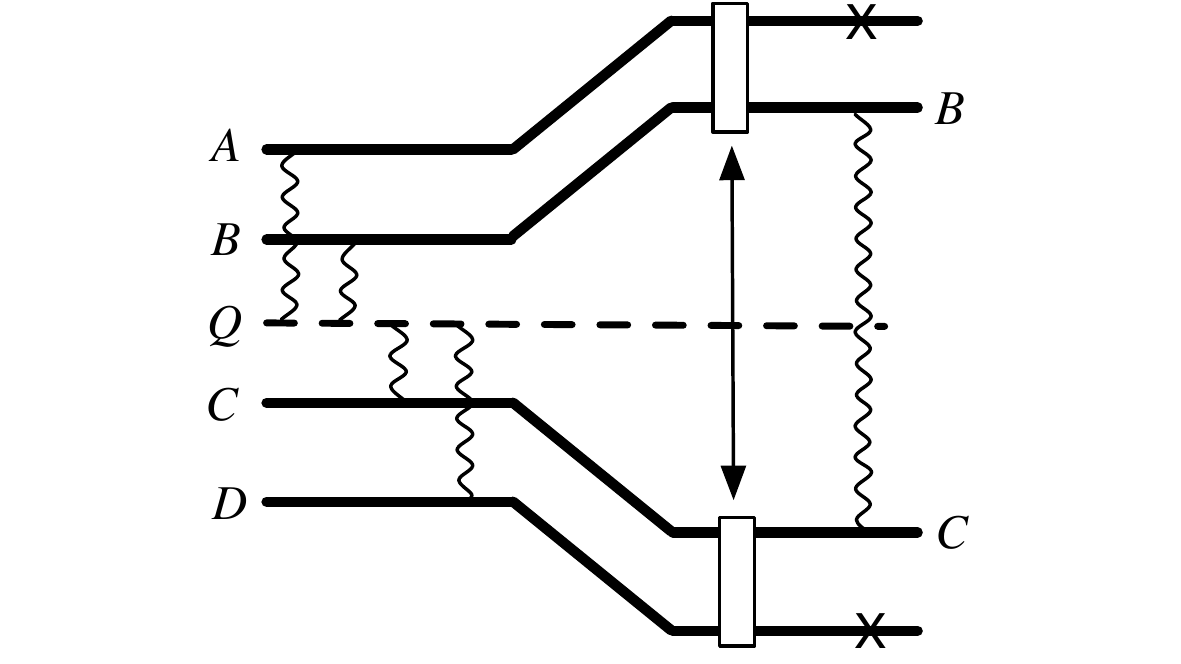}
   \caption{Remote entanglement between $B$ and $C$ is generated by first entangling (wavy lines) four systems $A$, $B$, $C$, and $D$ with a quantum state $Q$, moving the pairs $AB$ and $CD$ to remote locations, and then disentangling $A$, $D$, and $Q$ via local operations (rectangular boxes) and possibly communication (arrow), so that the entangled pair $BC$ is fully disentangled from the rest.}
   \label{fig:sketch}
\end{figure}

 The goal of our procedure is to generate entanglement between two systems $B$ and $C$ that have never interacted. First, we consecutively entangle four qubits $A$, $B$, $C$, and $D$ with a quantum system $Q$ to generate the joint state $ABCD$. Then, the qubit pairs $AB$ and $CD$ are sent to remote locations where, finally, $B$ and $C$ are disentangled locally to create the entangled pair $BC$ in product with $AD$ (see Fig.~\ref{fig:sketch}).

 The protocol sketched in Fig.~\ref{fig:sketch} appears formally similar to entanglement purification protocols~\cite{BDSW1996}, whose purpose is to distill pure entangled states from entangled mixed states. Such protocols are iterative, and act on $m$ copies of the quantum state to distill $n<m$ pure entangled states that can then be used as a resource for other protocols, and either require one-way or two-way communication between the remote destinations. The purpose of the present protocol is different: we prepare the initial quantum state in a defined manner so that remote bi-partite entanglement can be ``activated'' on demand, using one-way communication in general, and without any communication in special cases.  

 To enable this protocol, four ancillary qubits $A$, $B$, $C$, and $D$ are first encoded via a sequence of measurements~\cite{GlickAdami2017} of a quantum system $Q$. These measurements are equivalent to the unitary entangling operations implemented in weak measurements~\cite{Aharonov-weak-1988,Ritchie1991,Hosten2008,Lundeenetal2011,LundeenBamber2012,Dressel2014}, but the interactions considered here are strong (see, e.g.,~\cite{Nakamura2012,Vallone2016}). As we will see, the degree of entanglement of the resulting state $BC$ will depend on the relative angles between measurement bases.

 The density matrix of the initial qubit quantum system is taken to be proportional to the identity matrix, $\rho_Q = \frac12 \, \mathbbm{1}$, so that it is a maximum entropy state. The four consecutive entangling operations between the quantum system and the ancillary qubits lead to the total wave function~\cite{CerfAdami1998,GlickAdami2017},
 \begin{equation}\label{qrabcd-ancilla-basis}
      |QRABCD\ra = \frac{1}{\sqrt 2} \sum_{i j k \ell} U_{ij} \, U'_{jk} \, U''_{k\ell} \, |\ell i \, ijk\ell\ra,
 \end{equation}
 where the initial mixed state of $Q$ has been purified with a reference system $R$ and each system is of dimension two so that entropies are in units of bits. The measured observables of $Q$ are related through the matrix elements of $U$, $U'$, and $U''$ that specify the measurement bases. That is, the eigenbasis of the first observable (indexed by $i$) measured using qubit $A$ is rotated relative to the eigenbasis of the second observable (indexed by $j$) measured with $B$ according to $|i\ra = \sum_j U_{ij} \, |j\ra$. Similar expressions hold for $U'$, where the third observable (indexed by $k$)  is measured using $C$, and for $U''$, where the fourth (indexed by $\ell$) is measured with $D$. Since the measurements are strong, the final states of the ancillary qubits in~\eqref{qrabcd-ancilla-basis}, $|i\ra_A$, $|j\ra_B$, $|k\ra_C$, and $|\ell\ra_D$, are orthogonal. 
 
 Without loss of generality we can consider only measurements of observables corresponding to the $xz$ plane of the Bloch sphere, so that the transformation $U$ can be implemented with
 \begin{equation}
    U = 
    \begin{pmatrix}
	\cos\theta & -\sin\theta \\
	\sin\theta &  ~~~\cos\theta
    \end{pmatrix},
 \end{equation}
 for a rotation by an angle $\theta$, and similarly for $U'$ and $U''$ with angles $\theta'$ and $\theta''$, respectively. With this parametrization, a rotation by $\pi/4$ from, e.g., the eigenbasis of $\sigma_z$ yields the eigenbasis of $\sigma_x$.
 
 The encoding operation starts by consecutively entangling qubits $A$ and $B$ with $Q$, with a relative angle of $\theta$ between the first two observables, and sending them to Alice. Qubits $C$ and $D$ are subsequently entangled with $Q$, with a relative angle $\theta''$ between the last two observables, and are sent to Bob. As we will see, the relative angle $\theta'$ between the second and third observables can be left arbitrary implying that it is not necessary for Alice or Bob to know the measurement bases the other chose.  The resulting chain~\eqref{qrabcd-ancilla-basis} of ancillary qubits is coherent~\cite{GlickAdami2017}, and it is this fact that will allow us to deterministically create remote bi-partite entanglement. 
 
 The chain's joint state can be written in terms of a new joint basis for $B$ and $C$ so that, tracing~\eqref{qrabcd-ancilla-basis} over $Q$ and $R$, it appears as
 \begin{equation}\label{abcd}
      \rho_{ABCD} = \frac12 \sum_{i \ell} p_{i\ell} \,  |i\ra\la i| \otimes|\phi_{i \ell} \ra\la \phi_{i \ell}| \otimes |\ell\ra \la \ell |.
 \end{equation} 
 The four non-orthogonal joint states of $B$ and $C$, 
 \begin{equation}\label{conditional-states}
    \epsilon_{i\ell} \, |\phi_{i \ell}\ra = \sum_{jk} U_{ij} \, U'_{jk} \, U''_{k\ell} \, |jk\ra,
 \end{equation}
 are normalized according to
 \begin{equation}\label{normalization}
    p_{i\ell} = |\epsilon_{i\ell}|^2  = \sum_{jk} |U_{ij}|^2 \, |U'_{jk}|^2 \, |U''_{k\ell}|^2.
 \end{equation}
 
 To generate remote entanglement, Alice (who holds the pair $AB$) and Bob (in possession of $CD$) must perform a set of conditional unitary operations that are consistent with the choice of encoding parameters ($\theta$, $\theta'$, and $\theta''$), so that after the operations they share an entangled state of $B$ and $C$ that is in a product state with the rest of the system. As local unitary operations alone cannot change the entanglement of a state~\cite{NielsenChuang_Book2010},
 remote entanglement can only be generated in this protocol by local operations on $B$ and $C$ when the entanglement entropies of each state $|\phi_{i\ell}\ra$ in~\eqref{conditional-states} are the same. The structure of these operations, and the restrictions that must be put on the angles in order to create a maximally entangled pair, will be examined next.
      
 The first protocol we describe in Sec.~\ref{sec:LOCC} is the most general and can be implemented when Alice chooses the relative angle $\theta = \pi/4$ (Bob could, equivalently, pick $\theta'' = \pi/4$). We will see in Sec.~\ref{sec:LOCC} that, in this case, classical communication of the states of qubits $A$ and $D$ is required, but (of course) {\em not} of the states of $B$ and $C$. Furthermore, the final degree of entanglement between $B$ and $C$ will depend only on Bob's relative measurement angle $\theta''$, and not on $\theta'$. 
 
 In Sec.~\ref{sec:LO}, we consider a special case of the general scheme where both the first and third relative angles are set to $\theta = \theta'' = \pi/4$. We will find that communication about the states of $A$ and $D$ is no longer required and a maximally entangled state is deterministically extracted by operations applied locally to the two pairs of qubits. Interestingly, in the general and special cases, Alice and Bob do not need to communicate their measurement bases since the operations they implement are independent of all relative measurement angles, including the relative angle $\theta'$ between the measurements with $B$ and $C$.

 \section{Protocol with communication} 
 \label{sec:LOCC}
 
 The general protocol described here requires classical information to be communicated between Alice and Bob to generate remote entanglement. Using local operations on $B$ and $C$ in addition to classical communication about the states of $A$ and $D$, a new joint state of $B$ and $C$ is generated that is entangled according to Bob's relative measurement angle $\theta''$.
 
 To encode the ancillary qubits, Alice selects the relative angle $\theta = \pi/4$ while Bob's angle $\theta''$ is left arbitrary. In this case, it is straightforward to show that the conditional joint states~\eqref{conditional-states} of qubits $B$ and $C$, which are functions of the relative angles $\theta'$ and $\theta''$,
 can each be written in terms of local operations on the state $|\phi_{00}\ra$,
 \begin{equation}\label{phi-i-l-LOCC-2}
      |\phi_{i\ell}\ra = V^{(i\ell)\dagger}\, |\phi_{00}\ra,
 \end{equation}
 where
 \begin{equation}\label{V-i-l-LOCC}
    V^{(i\ell)\dagger} = Z^{i+\ell}X^\ell \otimes X^\ell 
 \end{equation}   
 are conditional unitary operators on qubits $B$ and $C$. Here, $Z$ and $X$ are Pauli operators and the sum $(i+\ell)$ is modulo two. We write the $i = \ell = 0$ state, 
 \begin{equation}\label{phi00-2}
      |\phi_{00}\ra =  -\sin\theta' \, |\widetilde{\beta}_{01} \ra + \cos\theta' \, |\widetilde{\beta}_{10} \ra,
 \end{equation}
 in terms of the generalized Bell basis,
 \begin{equation}
 \begin{split}
      |\widetilde{\beta}_{00}\ra & = \sin\theta'' |00\ra + \cos\theta'' |11\ra,\\
      |\widetilde{\beta}_{01}\ra & = \sin\theta'' |01\ra + \cos\theta'' |10\ra,\\
      |\widetilde{\beta}_{10}\ra & = \cos\theta'' |00\ra - \sin\theta'' |11\ra,\\
      |\widetilde{\beta}_{11}\ra & = \cos\theta'' |01\ra - \sin\theta'' |10\ra.
 \end{split}
 \end{equation}
 
 Using~\eqref{phi-i-l-LOCC-2}, the joint density matrix~\eqref{abcd} of all four qubits then becomes
 \begin{equation}\label{abcd-LOCC}
      \rho_{ABCD} = \frac14  \sum_{i \ell} |i\ra \la i| \otimes V^{(i\ell)\dagger}|\phi_{00} \ra \la \phi_{00}|\, V^{(i\ell)} \otimes |\ell\ra \la \ell |.
 \end{equation}
 We can see from~\eqref{abcd-LOCC} that if Alice and Bob apply the Hermitian conjugate of the operators~\eqref{V-i-l-LOCC}, then qubits $B$ and $C$ will be left in the entangled state~\eqref{phi00-2} that is in a product state with the rest of the system. This is the essence of the remote entanglement generation scheme. 
 
 The operation that disentangles the qubits $B$ and $C$ from the rest of the system takes the form
 \begin{equation}\label{disentangle-LOCC}
      V = \sum_{i \ell} |i\ra\la i| \otimes V^{(i\ell)}\otimes |\ell\ra\la\ell|,
 \end{equation}
 where the unitary operators $V^{(i\ell)}$ were defined in~\eqref{V-i-l-LOCC}. This expression does not completely factor into two separate operations on the pairs of qubits $AB$ and $CD$, meaning that classical communication is necessary between Alice and Bob in order to implement~\eqref{disentangle-LOCC}. Indeed, it is clear from~\eqref{V-i-l-LOCC} that Alice must know the state $\ell$ of Bob's qubit $D$ before performing her operations on $A$ and $B$, while Bob does not need to know the state $i$ of Alice's qubit $A$ (one-way communication). Despite the communication requirement, the operations~\eqref{V-i-l-LOCC} are independent of the angles $\theta'$ and $\theta''$ so that Alice and Bob do not need to know each other's measurement bases. 

 After applying~\eqref{disentangle-LOCC} to the state~\eqref{abcd-LOCC},
 \begin{equation}\label{disentangled}
      V\rho_{ABCD} V^\dagger = \frac12 \, \mathbbm{1} \otimes |\phi_{00}\ra\la \phi_{00}| \otimes \frac12 \, \mathbbm{1}, 
 \end{equation}
 qubits $B$ and $C$ are left in the pure state~\eqref{phi00-2}, which is in a product state with $A$ and $D$. In fact, $B$ and $C$ are disentangled from the {\em entire} system due to the symmetry of the underlying state~\eqref{qrabcd-ancilla-basis}.

 The degree of entanglement of the final state~\eqref{phi00-2} can be computed from the entanglement entropy, $S_E$. This quantity characterizes the entanglement of a bipartite pure state, and is the same for each state~\eqref{phi-i-l-LOCC-2} since local operations alone do not change the degree of entanglement. The entanglement entropy is computed from the von Neumann entropy of one of the subsystems, e.g., $ \rho_B^{(i\ell)} = {\rm Tr}_C (|\phi_{i\ell}\ra\la\phi_{i\ell}|)$, and turns out to be independent of the angle $\theta'$, 
 \begin{equation}\label{entanglement-entropy}
 \begin{split}
      S_E & = S\big( \rho_B^{(i\ell)} \big) \\ 
      & = -\cos^2\!\theta'' \log \cos^2\!\theta'' \!-\sin^2\!\theta'' \log \sin^2\!\theta''\!.
 \end{split}
 \end{equation}
 Evidently, the conditional states~\eqref{phi-i-l-LOCC-2} are uncorrelated when $\theta'' = 0$ ($S_E = 0$) and are fully entangled when $\theta'' = \pi/4$ ($S_E = 1$). This second case is the one considered in Sec.~\ref{sec:LO} where the entanglement generated is maximal. Thus, Bob can control the entanglement of the final shared joint state~\eqref{phi00-2} of $B$ and $C$ by choosing a particular relative angle $\theta''$.

 \subsection{Conditions for entanglement generation}
 
 In the most general scenario where all three relative angles are left arbitrary, the entanglement entropy of~\eqref{conditional-states} must be a constant for all $i,\ell$ for local operations to successfully disentangle the joint state of qubits $B$ and $C$ from the rest of the system. In other words, it is necessary for the Schmidt coefficients of the Schmidt decomposition of each state~\eqref{conditional-states} to be the same. This, however, does not guarantee that the resulting joint state of $B$ and $C$ at the end of the protocol will be {\em entangled}. For instance, at $\theta = 0$ or $\theta''=0$, the entanglement entropies of~\eqref{conditional-states} are indeed all the same, but vanish, so that the resulting joint state of $B$ and $C$ is completely uncorrelated and no shared entanglement is created.
 
 Interestingly, the shared entropy of $A$ and $D$ vanishes,
 \begin{equation}\label{condition}
     S(A:D) = S(A) + S(D) - S(AD) = 0,
 \end{equation} 
 only when nonzero entanglement between qubits $B$ and $C$ is successfully generated, and is otherwise positive. Given the correlated structure of the coefficients $p_{i\ell}/2$ [see~\eqref{normalization}] in the density matrix $\rho_{AD}$ (found by tracing~\eqref{abcd} over $B$ and $C$), a vanishing mutual entropy can only occur if at least one of the three angles is $\pi/4$ so that $p_{i\ell}/2 = 1/4$. In turn, this corresponds precisely to a constant and {\em nonzero} entanglement entropy. Thus, the necessary and sufficient condition for the entanglement generation scheme described here is simply~\eqref{condition}.

 \begin{figure}[t]
  \centering 
  \includegraphics[width=\linewidth]{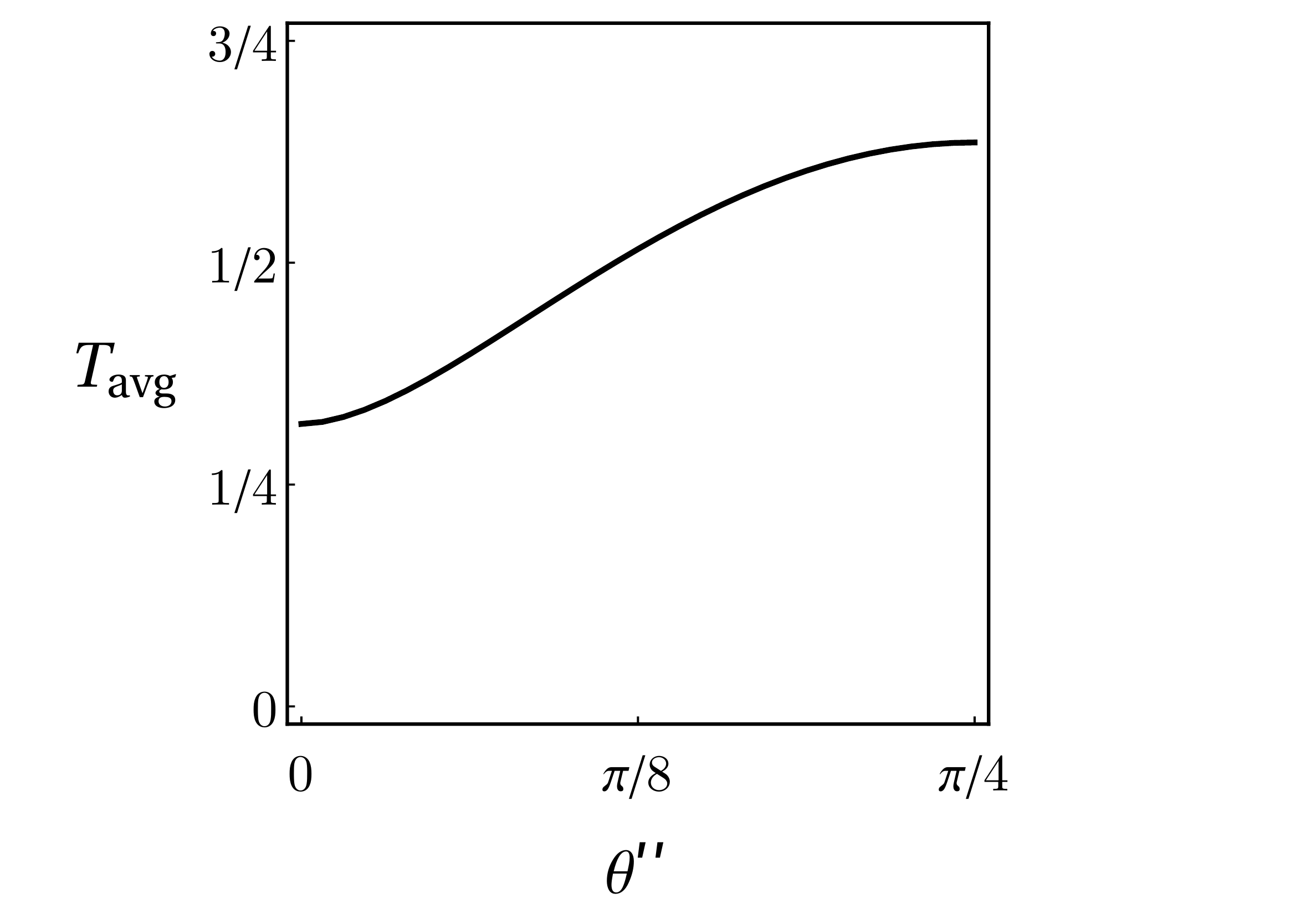}
  \caption{The trace distance, $T$, between~\eqref{abcd} and~\eqref{abcd-classical} averaged over all intermediate measurement angles $\theta'$, as a function of $\theta''$. Here, $\theta = \pi/4$.}
  \label{fig:distance-2}
 \end{figure}

 \subsection{Reliability}
 
 The encoding in this protocol serves to protect the conditional coherence of qubits $B$ and $C$ by correlating them with $A$ and $D$. These correlations make it possible to generate shared entanglement between $B$ and $C$ without requiring joint measurements or even, as in the case of the protocol we discuss in Sec.~\ref{sec:LO}, classical communication. 
 
 To quantify the effect of such correlations, we study the coherence of the joint state~\eqref{abcd} of all four ancillary qubits before the disentangling operation~\eqref{disentangle-LOCC} is applied. The closer the state~\eqref{abcd} is to its classical counterpart, the more ``robust'' it may be to decoherence during the protocol. We determine how close $\rho_{ABCD}$ is to the classical (completely decoherent) version denoted by $\sigma_{ABCD}$, which has only diagonal elements, by computing the trace distance between~\eqref{abcd} and
 \begin{equation}\label{abcd-classical}
      \sigma_{ABCD} = \frac12 \sum_{i j k \ell} |U_{ij}|^2 \, |U'_{jk}|^2 \, |U''_{k\ell}|^2 \, |ijk\ell\ra\la ijk\ell|.
 \end{equation}
 The trace distance~\cite{NielsenChuang_Book2010}, $T = \frac12 {\rm Tr}(|\rho - \sigma|)$, between two states $\rho$ and $\sigma$ is averaged over all intermediate angles $\theta'$ and plotted in Fig.~\ref{fig:distance-2} for $\theta = \pi/4$. The average trace distance $T_{\rm avg}$ remains less than one, suggesting that the protocol may be more robust to decoherence when qubits $B$ and $C$ are correlated with $A$ and $D$. As Bob's relative angle $\theta''$ increases, $T_{\rm avg}$ increases and the resulting joint state~\eqref{phi00-2} of $B$ and $C$ becomes more entangled.

 \section{Protocol without communication}
 \label{sec:LO}
 
 The protocol described in Sec.~\ref{sec:LOCC} encoded the ancillary qubits with the relative angle $\theta = \pi/4$, while $\theta'$ and $\theta''$ were left arbitrary. Here, we look at a special case of that general scheme where now the first and third relative angles are set to $\theta = \theta'' = \pi/4$. We will see that the operators that replace~\eqref{V-i-l-LOCC} completely factorize in this case, which eliminates the communication requirement between Alice and Bob. 
 
 Setting $\theta = \theta'' = \pi/4$, the conditional operators on $B$ and $C$ that appear in the density matrix~\eqref{abcd-LOCC} are now given by
 \begin{equation}\label{V-i-l-LO}
    \widetilde{V}^{(i\ell)\dagger} = Z^i \otimes (-Z)^\ell.
 \end{equation}   
 There are two important features of~\eqref{V-i-l-LO}. As before, the set of operators on $B$ and $C$ do not depend on the intermediate angle $\theta'$, but now they are completely factorized. That is, the operator $Z^i$, with only the index $i$, is applied to qubit $B$, while $(-Z)^\ell$, with only the index $\ell$, acts on qubit $C$. As a consequence, Alice and Bob do not need to know the state of the other's qubit in order to implement the operation
 \begin{equation}\label{controlled-U}
      \widetilde{V} \!=\! \left[ \sum_i |i\ra \la i| \otimes Z^i \right] \!\otimes\! \left[ \sum_\ell \left( -Z \right)^\ell \otimes |\ell\ra\la\ell| \right]\!.
 \end{equation}
 It is clear from~\eqref{controlled-U} that if Alice and Bob each apply a controlled-phase gate to their pair of qubits (with the controls on $A$ and $D$), the resulting state will be~\eqref{disentangled}, where $|\phi_{00}\ra$ is now given by
 \begin{equation}\label{phi00}
      |\phi_{00}\ra =  -\sin\theta' \, |\beta_{01} \ra + \cos\theta' \, |\beta_{10} \ra.
 \end{equation}
 Here, we used the standard Bell basis,
 \begin{equation}
      |\beta_{zx}\ra = (\mathbbm{1} \otimes X^x Z^z) \, |\beta_{00}\ra,
 \end{equation}
 where $|\beta_{00}\ra = |\Phi^+\ra$ is the usual Bell state. The remaining three conditional joint states of $B$ and $C$ are obtained by applying~\eqref{V-i-l-LO} to the $i = \ell = 0$ state~\eqref{phi00}.
 
 Thus, with operations applied locally to their pairs of qubits, Alice and Bob extract a joint pure state of $B$ and $C$ that is in a product state with the rest of the system. We emphasize that this does not require any classical communication between Alice and Bob, and that afterwards they share one half each of the entangled state~\eqref{phi00}. It is easy to show that the state~\eqref{phi00} is maximally entangled regardless of the angle $\theta'$ (its entanglement entropy is equal to one). Since the operators~\eqref{V-i-l-LO} on $B$ and $C$ are independent of the angle $\theta'$, the protocol can be used even when Alice and Bob do not know each other's measurement bases. 

  \begin{figure}[t]
  \centering 
  \includegraphics[width=\linewidth]{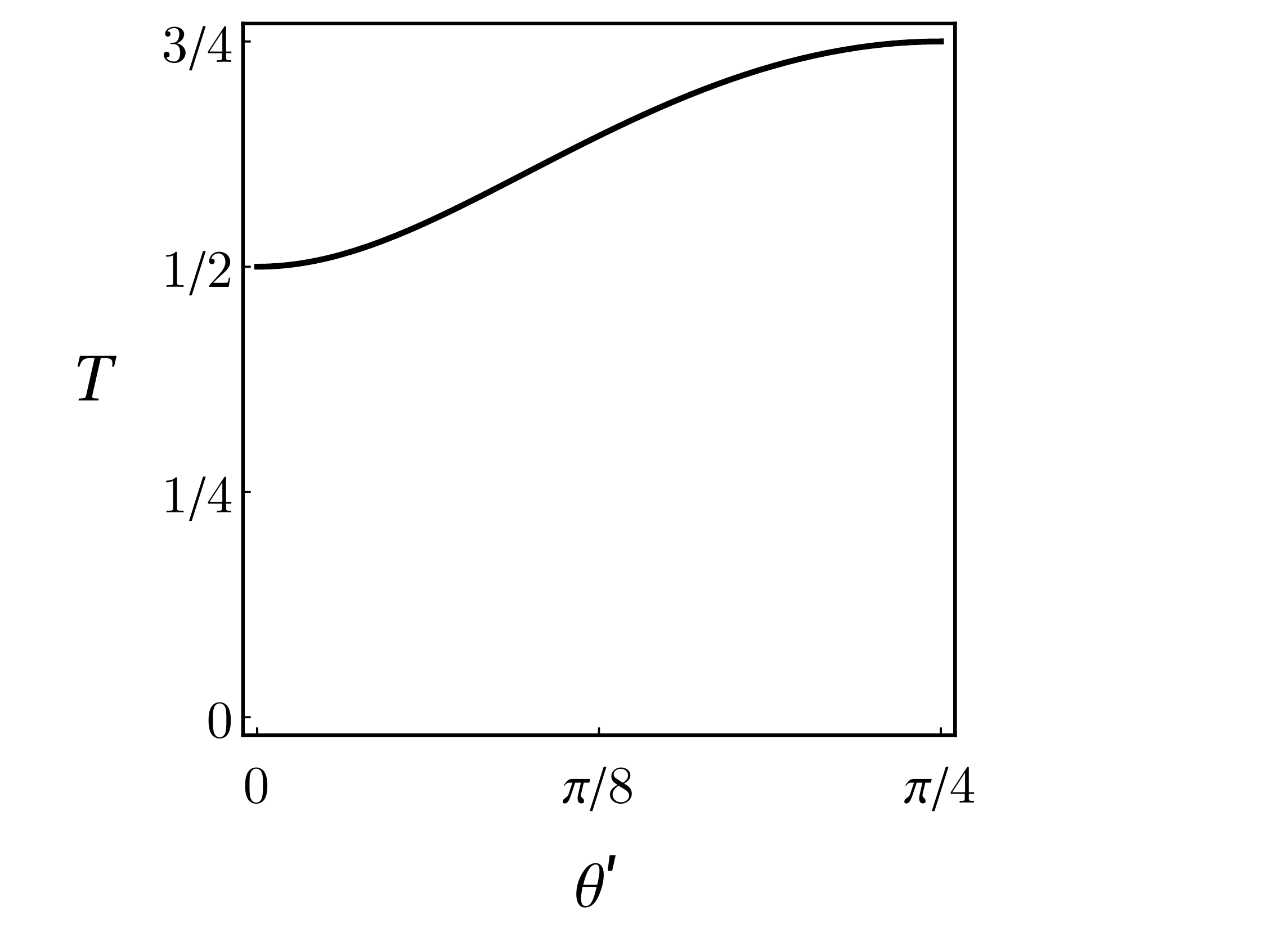}
  \caption{The trace distance, $T$, between~\eqref{abcd} and~\eqref{abcd-classical} as a function of the intermediate angle $\theta'$. Here, both Alice's and Bob's relative measurement angle is $\theta = \theta'' = \pi/4$.}
  \label{fig:distance}
  \end{figure}
    
 For the complete entanglement generation scheme without communication, refer to Fig.~\ref{fig:sketch}. In this protocol, the state of the four ancillary qubits after the encoding operations (the first four wavy lines) is given by~\eqref{abcd} with $\theta = \theta'' = \pi/4$, the rectangular boxes represent the controlled-phase gates in~\eqref{controlled-U}, and the arrow indicating classical communication is removed. The final state generated in the protocol corresponds to the maximally entangled state~\eqref{phi00}. We plot in Fig.~\ref{fig:distance} the trace distance between~\eqref{abcd} and~\eqref{abcd-classical} with $\theta = \theta'' = \pi/4$ as a function of the intermediate angle $\theta'$ and show that it remains less than one.

\section{Discussion}

 We described two methods for generating shared entanglement between remote parties that have never interacted in the past. Both techniques are deterministic, based on a simple encoding scheme, and do not require the joint measurements used in entanglement swapping. The general scheme requires one-way classical communication between Alice and Bob, and the degree of entanglement extracted can be tuned by the choice of encoding parameters (the set of relative measurement angles). A particular encoding of the initial state makes this protocol work even in the absence of communication. In that case, operations applied locally to the qubit pairs are sufficient to deterministically generate a maximally entangled state.
 
 We should point out that, given the nature of the state-preparation protocol, it is not necessary for the quantum states $A$, $B$, $C$, and $D$ to be co-located with the initial quantum state $Q$ when state preparation occurs, and then sent out to remote locations. Instead, we can imagine that the state $Q$ is first sent to Alice at a remote location, who measures $Q$ first with $A$ and then $B$, then sends $Q$ on to Bob at another location where he measures the same quantum system using $C$ and then $D$. 
 
 In this work, we have not discussed the effect of noise on the present protocol (for example, due to entanglement with uncontrolled degrees of freedom), which would undoubtedly result in a success rate smaller than one. In particular, it is likely that entanglement is better protected from decoherence before disentanglement (akin to an error-correcting code), so that ideally the parties would wait to ``activate'' the entanglement until just before it is needed. 
 
 That maximum entanglement can be generated even when Alice and Bob do not know each other's measurement bases (the disentangling operations are independent of the relative angle $\theta'$) could make this a useful scheme for quantum key distribution protocols. In this case, it would be useful if Alice and Bob could independently verify the entanglement they share (see, e.g.,~\cite{Enk2007}). Because $S(A:D)=0$ before and after the entanglement between $B$ and $C$ is created, qubits $A$ and $D$ are unlikely to be helpful for this purpose. Future work should be able to throw light on this and other issues including the effectiveness of the protocol in the presence of noise or an eavesdropper.

\bibliographystyle{unsrtnat} 
\bibliography{bibliography}

\end{document}